\documentclass[12pt]{article}
\usepackage{times}
\usepackage[dvips]{graphicx}
\usepackage[english]{babel}
\usepackage{amsmath}
\usepackage{amssymb}

\newenvironment{sciabstract}{%                                                 
\begin{quote} \bf}
{\end{quote}}

\newcounter{lastnote}

\title{Pairing Gap and In-Gap Excitations in Trapped Fermionic Superfluids}
\author{J. Kinnunen, M. Rodr\'{\i}guez, and P. T\"orm\"a$^\ast$ \\
\\
\normalsize{Department of Physics, NanoScience Center,}\\
\normalsize{P.O.Box 35, FIN-40014 University of Jyv\"askyl\"a, Finland}\\
\\
\normalsize{$^\ast$To whom correspondence should be addressed; E-mail: 
paivi.torma@phys.jyu.fi.}
}

\date{}

\begin{document}

\baselineskip24pt
\maketitle

\begin{sciabstract}
We consider trapped atomic Fermi gases with Feshbach-resonance enhanced interactions
in pseudogap and superfluid temperatures. We calculate the spectrum of 
RF(or laser)-excitations for transitions that transfer atoms out of the superfluid 
state. The spectrum displays the pairing gap and also the contribution of unpaired
atoms, i.e.\ in-gap excitations. The results support the conclusion 
that a superfluid, where pairing is a many-body effect, was observed in recent 
experiments on RF spectroscopy of the pairing gap.  
\end{sciabstract}

\newpage

Fermionic superfluidity and superconductivity appear in several systems in nature 
such as metals, cuprates and helium. In the limit of weak interparticle interaction, 
the Bardeen-Schrieffer-Cooper (BCS) theory of superconductivity has been successful 
in explaining the observed phenomena as a Bose-Einstein Condensation (BEC) of weakly 
bound momentum-space pairs. In the limit of strong interactions, spatially small, 
strongly bound pairs are formed and undergo BEC. The intriguing question about the 
nature of the crossover from BCS pairing to BEC of dimers was theoretically addressed in 
1980 ({\it 1, 2}) and is closely related to uncovering the nature of 
high-temperature superconductivity. Trapped fermionic atoms offer a system where 
the crossover can be scanned by tuning the inter-particle scattering length 
using Feshbach resonances ({\it 3-7}). At the crossover 
region, the scattering length diverges and a universal behaviour, independent of any 
length scale, is expected. The system is also genuinely mesoscopic due to the 
trapping potential for the atoms. Here we consider spectroscopic signatures of 
pairing in these systems at the onset of the superfluid transition and show that 
the mesoscopic nature of the system leads to pronounced signatures from unpaired 
atoms which can also be understood as in-gap excitations. The results are 
in agreement with the experimental results in ({\it 8}). 

The single-particle excitation spectrum of a fermionic superfluid is expected to 
show an energy gap. A spectroscopic method for observing the excitation gap in 
atomic Fermi gases has been proposed ({\it 9-11}). RF-spectroscopy 
has been used for observing mean fields ({\it 12, 13}) and, very recently, the 
excitation gap ({\it 8}). Laser- or RF-fields are used for transferring atoms 
out of the superfluid state to a normal one. The superfluid state originates from the %%@
pairing of 
atoms in two different internal states, say $|1\rangle$ and $|2\rangle$. The field 
drives a transition from $|2\rangle$ to a third state $|3\rangle$; atoms in state 
$|3\rangle$ are not paired i.e.\ they are in the normal state. The idea is closely 
related to observing the superconductor - normal metal current in metals and, 
similarly, it reflects the density of states and displays the excitation gap. 
Only, in this case, the superfluid-normal interface is realized by internal states of the %%@
atom, 
not by a spatial boundary. The response, in the case of atoms, is qualitatively 
different from that of metals due to the exact momentum conservation in atomic transitions 
driven by homogeneous fields. Here, we calculate the response of this process, 
that is, the spectrum as a function of the detuning of the RF-field, taking into account 
the mesoscopic nature of the sample i.e.\ the trapping potential. This leads to 
pronounced signatures which can be utilized in confirming the onset of the 
excitation gap and the superfluid transition.    

In the high-$T_c$ region of the BEC-BCS crossover, the BCS theory, in its 
simplest form, is expected to be incapable of describing the effects of strong 
interactions, such as the formation of a pseudogap. In atomic Fermi gases, the 
vicinity of the Feshbach resonance is associated with strong interactions, 
and preformed pairs causing a pseudogap may exist even above the critical 
temperature. The excitation gap, therefore, has contributions both from the 
superfluid gap ($\Delta_\mathrm{sf}$) and the pseudogap ($\Delta_\mathrm{pg}$).
The many-body state is affected also by the existence of the molecular bound 
state which actually causes the Feshbach resonance phenomenon. These issues 
are considered in recent theory work on resonance superfluidity 
({\it 14-17}). We use such an approach for 
calculating the equilibrium state of the system ({\it 18}). 

The interaction with the (RF/laser) field is introduced as a perturbation 
and the response is calculated to the second order in the perturbation Hamiltonian. 
This corresponds to a Fermi Golden rule -type of derivation of the spectrum
and allows a treatment of the complex many-body state with reasonable accuracy. 
The Hamiltonian $H_T$, describing the effect of the field, couples the 
states $|2\rangle$ and $|3\rangle$ ({\it 18}). The offset from the resonance of 
the transition between $|2\rangle$ and $|3\rangle$ is given by the RF-field detuning
$\delta = E_{RF} - (E_3 - E_2)$, where $E_{RF}$ and $E_3$, $E_2$ are the energies of the 
RF-photon and of the states $|3\rangle$ and $|2\rangle$, respectively.
The spectrum is obtained from the response 
$I (\delta) = \langle \dot N_\mathrm{3} \rangle$,
where $N_\mathrm{3}$ is the number of atoms in state $|3\rangle$, by 
neglecting terms of higher than second order in $H_\mathrm{T}$ in the 
derivation ({\it 18}). 
In the case of metals, such quantity would give the current $I(V)$, where $V$ is voltage, 
over the superconductor - normal metal tunneling junction. 

\begin{figure}[p]
  \centering
  \includegraphics[width=10cm]{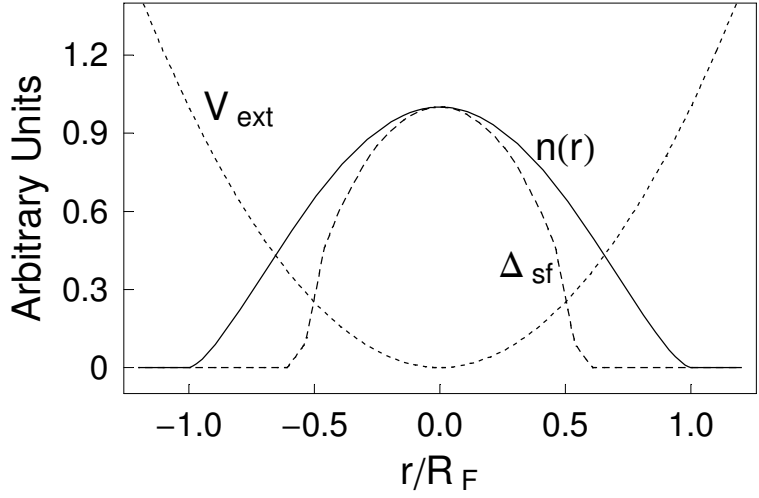}
  \caption{The superfluid gap and the atom density as functions of position at 
temperature $T=0.2\,T_\mathrm{F}$. Resonance superfluidity theory incorporating 
a pseudogap, together with Thomas-Fermi distribution in the local density 
approximation, is used. Only the atoms in the middle of the trap are condensed 
while the atoms closer to the borders are either free or in the pseudogap regime. 
The critical temperature in the middle of the trap is 
$T_\mathrm{c} \approx 0.3\,T_\mathrm{F}$.}
   \label{fig:scpos}
\end{figure}

\begin{figure}[p]
  \centering
  \includegraphics[width=10cm]{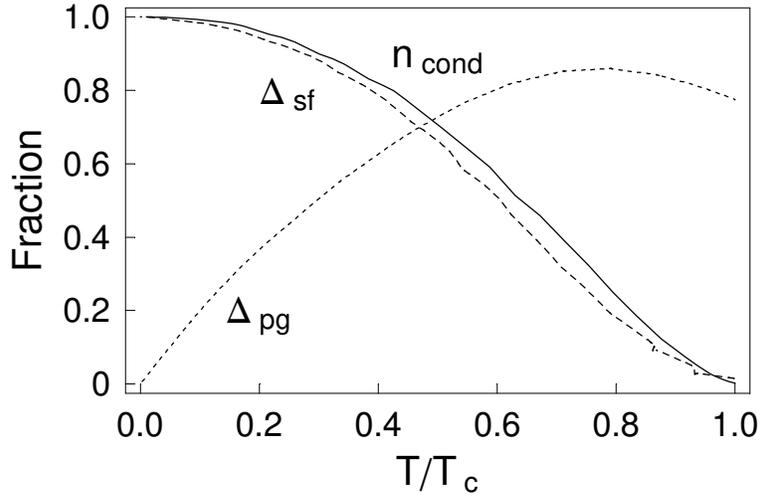}
  \caption{The mean superfluid gap ($\Delta_\mathrm{sf}$) and pseudogap 
($\Delta_\mathrm{pg}$) as functions of temperature. The fraction of condensed atoms
$n_\mathrm{cond}$ is defined as the fraction of atoms for which the temperature is 
below the local critical temperature. The temperature $T \approx 0.7\,T_\mathrm{c}$ 
corresponds to $T = 0.2\,T_\mathrm{F}$, showing that the superfluid gap distribution 
in Fig.~\ref{fig:scpos} corresponds to a condensate fraction of 
$n_\mathrm{cond} \approx 0.3$.}
   \label{fig:scmean}
\end{figure}

\begin{figure*}[p]
  \centering
  \includegraphics[width=\textwidth]{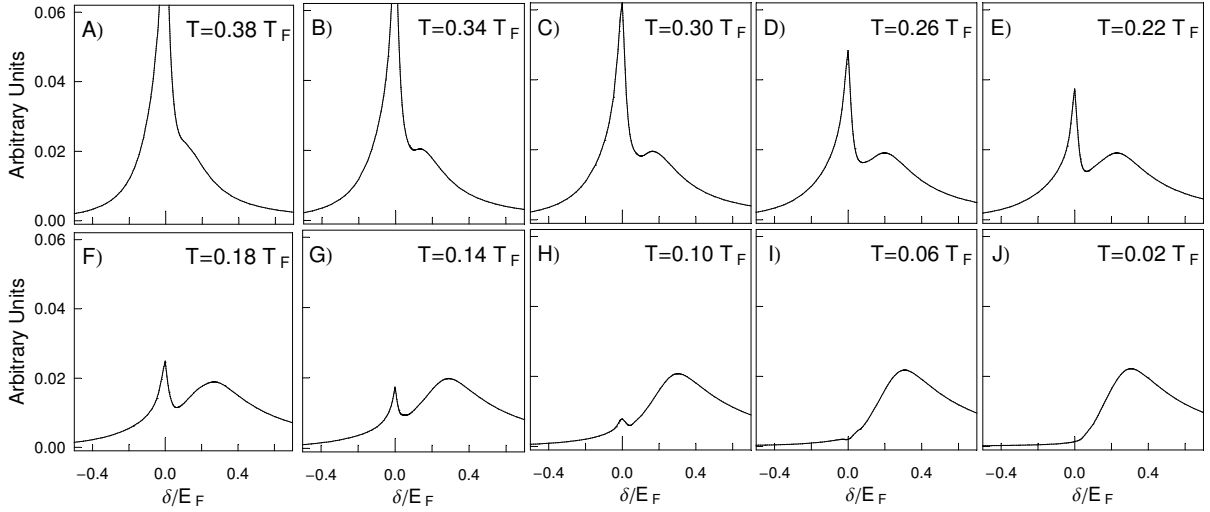}
  \caption{The spectra of the considered RF-transition as a function of the 
RF-field detuning $\delta$ for several temperatures. The peak at $\delta = 0$ is 
caused by free atoms. A peak shifted to the right from the zero gradually appears 
for lower temperatures, corresponding to paired atoms; the shift of the peak from 
the zero detuning gives the energy gap in the single particle excitation spectrum. 
The shift, that is, the gap grows with decreasing temperature. The plots show the 
disappearance of the free-atom peak when also the atoms at the borders of the trap 
enter the pseudogap regime and become paired. The critical temperature in the middle 
of the trap is $T_\mathrm{c} \approx 0.3\,T_\mathrm{F}$. At the temperature 
$T = 0.1\,T_\mathrm{F}$, more than 80\% of the atoms are condensed. The parameters 
used in the calculation correspond to the experiments in Fig.3 of ({\it 8}). 
The gas is in the unitarity limit, i.e.\ close vicinity of the Feshbach 
resonance, which is the expected high-$T_c$ regime for the system. 
   \label{fig:cur}} 
\end{figure*}

Trapped atomic gases have an inhomogeneous density distribution $n(\bf{r})$ and 
therefore a spatially varying superfluid order parameter is expected.  We treat 
the problem in the local density approximation, that is, we solve the 
equilibrium state by including $n(\bf{r})$ given by the Thomas-Fermi distribution 
as a position dependent parameter ({\it 18, 19}). 
Fig.~\ref{fig:scpos} presents the position dependence of the atom density and the
superfluid gap. This shows that only the atoms in the middle of the trap are 
condensed. Fig.~\ref{fig:scmean} shows the fraction of condensed atoms and the mean 
(averaged over $r$) superfluid gap and pseudogap as functions of temperature. 
The parameters used in calculating the results in Figs.~\ref{fig:scpos}--\ref{fig:cur} 
correspond to the experiments in Fig.~3 of ({\it 8}) and are given in ({\it 18}).

The spectra $I(\delta)$ at different temperatures are plotted in Fig.~\ref{fig:cur}. 
The peak at the zero detuning, 
$\delta = 0$, originates from free (unpaired) atoms. Another peak, shifted right 
from the zero, appears with decreasing temperature. The shift reflects the 
excitation gap, i.e.\ the energy needed for breaking a pair. The free-atom 
peak gradually vanishes when the temperature is lowered and also the atoms at 
the borders of the trap become paired. The disappearance of the free atom peak shows 
that the border atoms have reached the pseudogap regime ({\it 18}) and that the 
atoms in the middle of the trap are well below the superfluid transition 
temperature ({\it 20}). 
We have neglected the effect of the mean (Hartree-Fock) field energy shifts ({\it 18}), as %%@
they appear absent in the experiments ({\it 8, 13}). 

In a corresponding spatially homogeneous system, instead of the free-atom peak at 
zero detuning, a quasiparticle peak, shifted left from the zero, appears at high 
temperatures ({\it 11}). The shift is to the left, to the opposite 
direction than that of the pair-peak, because thermal quasiparticles of the superfluid 
already possess the excess gap energy, that is, energy is gained in the RF-transfer 
process ({\it 21}). As Fig.~\ref{fig:cur} shows, such quasiparticle peaks appearing 
in a homogenous system are now shadowed by trapping effects and the free-atom peak. 
The unpaired atoms in Fig.~\ref{fig:cur} can, however, be understood as in-gap 
excitations or quasiparticles. Instead of the local density approximation, inhomogenous 
superfluids can be described by the Bogoliubov - deGennes equations. Solving the 
equations in a trap geometry ({\it 10, 22}) results in in-gap excitations 
whose energies lie below the maximum (at the point of highest density) gap energy. 
The wave-functions of these excitations are located at the edges of the trap; they 
correspond to the free atoms at the borders of the trap in the local density treatment. 
The free atoms in Fig.~\ref{fig:cur} and observed in ({\it 8}) can thus be 
understood as in-gap excitations of an inhomogeneous superfluid.

The spectra in Fig.~\ref{fig:cur} are in excellent qualitative agreement with the 
experimental results in ({\it 8}). Also quantitatively they agree well 
with ({\it 8}) (c.f.\ Fig.3 in that article). The shift of the pair-peak, which 
gives the excitation gap, is at temperatures $T' \le 0.2 T_F$ about $0.2 E_F$ in 
({\it 8}) and $0.3 E_F$ for $T \le 0.1 T_F$ according our calculation. The widths of the %%@
peaks, 
which are determined by the gap, are about $0.3 E_F$ and $0.4 E_F$, respectively. 
The critical temperature at the center of the trap is in our case $T_c \sim 0.3 T_F$ 
which may be used to estimate that in ({\it 8}) it is $\sim$ 0.2-0.25 $T_F$.
The temperatures $T'$ in the experiment are determined in the BEC limit due to lack of 
precise thermometry in the unitarity limit. The adiabatic passage to the unitarity 
limit, where the spectra are actually measured, is expected to reduce the temperature 
due to entropy conservation so that $T < T'$ ({\it 23}). This is consistent with the %%@
observation 
that the pair-peak in Fig.~\ref{fig:cur} starts to appear at $T \sim 0.35 T_F$ 
and is clearly visible at $T \sim 0.2 T_F$, but in ({\it 8}) it appears 
and is clearly visible already at higher (BEC limit) temperatures of $T' \sim 0.75 T_F$ 
and $T' \sim 0.45 T_F$, respectively. The sensitivity of the free-atom (quasiparticle) 
peak to temperature and the possibility of direct comparison between theory and 
experiment may offer a route for developing a precise thermometry for the 
crossover region. 

We emphasize that those spectra in ({\it 8}) where the free-atom peak has 
disappeared correspond to the cases h) through j) in Fig.~\ref{fig:cur} where more 
than 80\% of the atoms are condensed. This indicates that the pairing observed 
at the lowest temperatures in ({\it 8}) corresponds to a superfluid. At higher 
temperatures, either a pseudogap or a combined effect of a superfluid gap and a 
pseudogap occurs. In summary, the results presented here support the conclusion 
that a superfluid, where pairing is a many-body effect, was observed 
in ({\it 8}). The mesoscopic nature of these novel Fermi superfluids shows up 
in an intriguing way. 

\subsection*{References and Notes} 
\begin{itemize}
\item[1.]
%\bibitem{Leggett}
A. \ J.\ Leggett, {\it Modern trends in the theory of Condensed Matter}, pg. 13 
(Springer-Verlag, 1980).  
%\bibitem{Nozi}
\item[2.]
P.\ Nozi\`eres, S.\ Schmitt-Rink, {\it J.\ Low Temp.\ Phys.\ } {\bf 59}, 195 (1985).
%\item{recentexp1}
\item[3.]
C.\ A.\ Regal , M. \ Greiner, D.\ S.\ Jin, {\it Phys.\ Rev.\ Lett.\ } \textbf{92}, 040403 %%@
(2004).
%\bibitem{exp2}
\item[4.]
M.\ Bartenstein \emph{et al.}, {\it Phys.\ Rev.\ Lett.\ } \textbf{92}, 120401 (2004).
%\bibitem{exp3}
\item[5.]
M.\ W.\ Zwierlein \emph{et al.}, {\it Phys.\ Rev.\ Lett.\ } \textbf{92}, 120403 (2004).  
%\bibitem{exp4}
\item[6.]
J.\ Kinast, S.\ L.\ Hemmer, M.\ E.\ Gehm, A.\ Turlapov, J.\ E.\ Thomas, 
{\it Phys.\ Rev.\ Lett.\ } \textbf{92}, 150402 (2004).
%\bibitem{exp5}
\item[7.]
M. Bartenstein \emph{et al.}, {\it Phys.\ Rev.\ Lett.\ } \textbf{92}, 203201 (2004).
%\bibitem{Grimm}
\item[8.]
C. Chin \emph{et al.}, Published online July 22 2004; 10.1126/science.1100818 (Science %%@
Express Reports).
%\bibitem{paivi}
\item[9.]
P.\ T\"orm\"a, P.\ Zoller, {\it Phys.\ Rev.\ Lett.\ } {\bf 85}, 487 (2000).
%\bibitem{Bruun}
\item[10.]
G.\ M.\ Bruun, P.\ T\"orm\"a, M.\ Rodr\'{\i}guez, P.\ Zoller,
{\it Phys.\ Rev.\ A} \textbf{64}, 033609 (2001).
%\bibitem{Kinnunen2004}
\item[11.]
J.\ Kinnunen, M.\ Rodr\'{\i}guez, P.\ T\"orm\"a, {\it Phys.\ Rev.\ Lett.\ } \textbf{92}, %%@
230403 (2004).
%\bibitem{rf1}
\item[12.]
C.\ A.\ Regal, D.\ S.\ Jin, {\it Phys.\ Rev.\ Lett.\ } \textbf{90}, 230404 (2003).
%\bibitem{rf2} 
\item[13.]
S.\ Gupta \emph{et al.}, {\it Science} \textbf{300}, 1723 (2003).
%\bibitem{ressup}
\item[14.]
M.\ Holland, S.\ J.\ J.\ M.\ F.\ Kokkelmans, M.\ L.\ Chiofalo, R.\ Walser, 
{\it Phys.\ Rev.\ Lett.\ } \textbf{87}, 120406 (2001).
%\bibitem{ressup2}
\item[15.]
E.\ Timmermans, K.\ Furuya, P.\ W.\ Milonni, A.\ K.\ Kerman,
{\it Phys.\ Lett.\ A} \textbf{285}, 228 (2001).
%E.\ Timmermans \emph{et al.}, Phys.\ Lett.\ A \textbf{285}, 228 (2001). 
%\bibitem{ressup3}
\item[16.]
Y.\ Ohashi, A.\ Griffin, {\it Phys.\ Rev.\ Lett.\ } \textbf{89}, 130402 (2002). 
%\bibitem{ressup4}
\item[17.]
J.\ Stajic \emph{et al.}, {\it Phys.\ Rev.\ A} \textbf{69}, 063610 (2004).
%\bibitem{online}
\item[18.]
Materials and methods are available as supporting material on {\it Science} Online.
\item[19.]
M.\ L.\ Chiofalo, S.\ J.\ J.\ M.\ F.\ Kokkelmans, J.\ N.\ Milstein, M.\ J.\ Holland,
{\it Phys.\ Rev.\ Lett.\ } \textbf{88}, 090402 (2002).
\item[20.]
The pronounced free-atom peak at the zero detuning clearly reflects the strong 
dependence of the order parameter on density and the trapping potential. In the case of 
a very smooth density dependence of the energy gap, one would expect simply a broadening 
of the pair-peak. Such type of behaviour was actually observed in the experiments 
probing mean fields ({\it 12,13}): there only one peak was observed and the effect 
of the trapping potential was a considerable broadening of the peak.
%\bibitem{note}
\item[21.]
Note that since not only energy but also momentum is exactly conserved in the 
transfer process, these particles cannot be transferred at zero detuning. This is 
in contrast to superconductor - normal metal junctions where thermal quasiparticle 
currents flow freely for below-gap voltages.
%\bibitem{Baranov2}
\item[22.]
M.\ A.\ Baranov, {\it JETP\ Lett.\ } \textbf{70}(6), 396 (1999).
\item[23.]
The precise relation between $T'$ and $T$ at the unitarity limit is not known but the 
analysis in case of a non-interacting Fermi gas ({\it 24}) supports this argument. 
%\bibitem{Carr}
\item[24.]
L.D.\ Carr, G.V.\ Shlyapnikov, Y.\ Castin, {\it Phys.\ Rev.\ Lett.\ } \textbf{92}, 150404 %%@
(2004).
%\bibitem{ack}
\item[25.]
We thank C. Chin and R. Grimm for a stimulating exchange of results and very useful 
discussions. We acknowledge the financial support from Academy of Finland 
(Project Nos.\ 53903, 205470) and the Emil Aaltonen foundation.
\end{itemize}

\noindent Supporting Online Material \\
www.sciencemag.org \\
Materials and Methods \\


\pagebreak

\subsection*{Supporting online material}

Feshbach resonances are a powerful tool to tune interactions in degenerate Fermi gases of 
atoms. They have been used for creation of molecules and molecular Bose-Einstein %%@
condensates 
(BEC) out of fermionic atoms ({\it 1-5}) and for exploring the 
crossover region between BEC of molecules and Bardeen-Schrieffer-Cooper (BCS) pairing of 
atoms ({\it 6-11}). At the crossover region, the scattering length diverges and universal 
behaviour independent of any length scale is expected ({\it 12-16}). 
This unitarity limit is predicted to be the high-$T_c$ region of the system ({\it 17-23}).

We consider a gas of atoms in two different internal states $|1\rangle$ and $|2\rangle$, 
corresponding to Fermion annihilation operators $c^{(1)}$ and $c^{(2)}$, respectively. The 
interaction between these states is enhanced by a Feshbach resonance. We denote the 
magnetic field detuning from the Feshbach resonance position by $\nu_0$. For $\nu_0 < 0$, 
the scattering length between the atoms is positive and two-body physics supports a 
molecular bound state, for which we introduce a bosonic annihilation operator $b$. At 
positive detunings $\nu_0$, the scattering length is negative and pairing which is a 
many-body effect is expected at low temperatures. Near the resonance, $\nu_0 \sim 0$, the 
scattering length diverges and the system is in the unitarity regime. 
We use the resonance superfluidity theory ({\it 17-21}) for calculating the equilibrium 
state of the system. The system is described by the Hamiltonian 
\begin{equation}
\begin{split}
  H&= \sum_{k,\sigma} \varepsilon^{\sigma}_{k} c^{\sigma\dagger}_{k}
    c^{\sigma}_{k} + \sum_{q}(E_q^{0}+\nu)b_{q}^\dagger b_{q} \\
  &+\sum_{q,k,k'}U(k,k')c_{q/2+k}^{(1)\dagger}c_{q/2-k}^{(2)\dagger}
    c_{q/2-k'}^{(2)}c_{q/2+k'}^{(1)} \\
  &+ \sum_{q,k}(g(k)b_q^\dagger c^{(1)}_{q/2-k}
    c^{(2)}_{q/2+k}+ h.c.).
\end{split}
\end{equation}
The interaction parameters $U$, $g$ and the Feshbach detuning $\nu$ are 
obtained from the bare parameters $U_0$, $g_0$ and $\nu_0$ by a renormalisation procedure 
({\it 18}). The momentum cutoff required by the summations and used in the 
renormalisation is chosen as $K_c = 25\,k_\mathrm{F}$, where $k_\mathrm{F}$ is
the Fermi momentum.
The energies $\varepsilon_k = k^2/2m$ and $E_q = q^2/2M$ are the kinetic energies
of a free fermion (with mass $m$) and a composite boson (mass $M = 2m$), here we have 
chosen $h/(2\pi) = 1$ where $h$ is Planck's constant. 
Equilibrium parameters such as the chemical potential $\mu$, pseudogap $\Delta_{pg}$ and 
the order parameter $\Delta_{sf}$ are solved self-consistently, following ({\it 24}). 
The total excitation gap is given by $\Delta^2 = \Delta_{sf}^2 + \Delta_{pg}^2$.
For further details see ({\it 25}).

The interaction with the (RF/laser) field is introduced as a perturbation and the response 
is calculated to second order in the perturbation Hamiltonian. The Hamiltonian describing 
the effect of the field is
\begin{equation}
\begin{split}
H_T&=\sum_k\frac{\delta}{2}\left(c_k^{(3)\dagger}c_k^{(3)}-c_k^{(2)\dagger}c_k^{(2)}
-b^\dagger_k b_k \right) \\
 &+\sum_{kl}
\left( M_{kl} c^{(2)\dagger}_k c_l^{(3)} + h.c. \right) +\sum_{klq}
\left( D_{qkl} b_q^\dagger c^{(1)}_k c_l^{(3)} + h.c. \right), 
\end{split}
\end{equation}
where $M_{kl}$ and $D_{qkl}$ are proportional to the Rabi frequency of the field and 
$\delta = E_{RF} - (E_3 - E_2)$ is the RF-field detuning, where $E_{RF}$ and $E_3$, $E_2$ 
are the energies of the RF-photon and of the states $|3\rangle$, $|2\rangle$, %%@
respectively.
We neglect the bosonic contribution in the perturbation Hamiltonian, assuming
that the number of composite bosons is small. The assumption is well-founded at least on %%@
the 
attractive side of the Feshbach resonance, where the Feshbach detuning is positive.
Inclusion of the bosonic current is straightforward, but on the repulsive side and at the 
resonance one should consider many-particle correlation functions to get the correct
asymptotic behaviour ({\it 26-28}). The spectrum is obtained from the response 
\( I (\delta) = \langle \dot N_\mathrm{3} \rangle = i \langle \left[ (H+H_T), 
N_\mathrm{3}\right] \rangle \),
where $N_\mathrm{3}$ is the number of atoms in state $|3\rangle$, by 
neglecting terms of higher than second order in $H_\mathrm{T}$. 
Applying Matsubara Green's functions techniques, $I(\delta)$ can be written as
\begin{equation}
  I(\delta) = 2 \sum_{kl} \left| M_{kl} \right|^2 \mathrm{Im}\left\{ 
\sum_{x_n^\mathrm{(2)}}
n_\mathrm{F} (x_n^\mathrm{(2)}) G_\mathrm{(3)}^\mathrm{ret} (m,x_n^\mathrm{(2)}-\delta) 
\underset{z=x_n^\mathrm{(2)}}{\mathrm{Res}} G_\mathrm{(2)} (n,z) + n_\mathrm{F} 
(\epsilon_m^\mathrm{(3)}) G_\mathrm{(2)}^\mathrm{adv} (n,\epsilon_m^\mathrm{(3)}+\delta 
)\right\}, \label{curr}
\end{equation}
where $x_n^{(2)}$ are the (imaginary) poles of the Green's functions $G_{(2)}$ and $n_F$ 
are the Fermi distribution functions.

We assume the three-dimensional Thomas-Fermi density distribution for the gas, 
where the density of atoms at distance $r$ from the middle of the trap is 
\begin{equation}
  n(r) = n(0)\left( 1-\left(\frac{r}{R_\mathrm{TF}}\right)^2\right)^{3/2}.
\end{equation}
Here $n(0)$ is the density in the middle of the trap and $R_\mathrm{TF}$ is the 
Thomas-Fermi radius, i.e.\ the size of the atom cloud. 
We treat the problem in the local density approximation, that is, we solve 
the equilibrium state including $n(r)$ as a position dependent parameter. 
Note that our analysis is independent of the symmetry of the 
cloud, i.e.\ ball as well as cigar or pancake shapes are described by the same analysis %%@
with scaled
coordinates. The use of the local density approximation is well grounded when the %%@
correlation length $\xi$
over which the atoms affect each other is much smaller than the trap size $l = %%@
\sqrt{\hslash/m\omega}$,
where $\omega$ is the trapping frequency ({\it 29-30}). For typical traps ({\it 31}), the %%@
radial frequency
is of the order of $10\,\mathrm{kHz}$ yielding the radial trap size of $l \approx %%@
20000\,a_\mathrm{0}$
while the axial frequencies are smaller by at least one order of magnitude.
Using the correlation length of the order of $\xi = O(1/k_F)$ ({\it 29}) gives, for 
Fermi energies of the order of $2\,\mu \mathrm{K}$, $\xi \approx 3000\,a_\mathrm{0}$ and 
the condition is well satisfied.  

The trap parameters are calculated for the maximum atom density of $10^{13}\,1/cm^3$ in
the center of the trap corresponding to the Fermi energy of $2\,\mu K$. We use a
background scattering length of the order $a_\mathrm{bg} = -2000\,a_0$ 
({\it 32}), where
$a_0$ is the Bohr's radius, corresponding to the two lowest substates of the electronic 
1$s^2$2$s$ ground state of $^6$Li. With this $a_\mathrm{bg}$ one obtains as the background 
interaction energy 
$U_0 = -0.5\,E_\mathrm{F}$. The boson-fermion coupling parameter is $g_0 \sim 10\,E_F$ in 
our calculations. The coupling $g_0$ cannot be directly obtained from experiments. It is 
defined in ({\it 17}) as
$g_0 = \sqrt{\Delta \mu_{Li} \Delta B U_0}$, where $\Delta \mu_{Li}$ is
the magnetic moment difference for between the Feshbach state and the continuum state and
$\Delta B$ is the width of the Feshbach resonance. The RF-spectra are not sensitive to the 
exact value of $g_0$ but it affects the fraction of molecules in the system. In order to 
describe the system close to the resonance, we choose the magnetic field detuning in the 
unitarity limit, $\nu_0 = 1\,E_\mathrm{F}$ ({\it 33}). The state $|3\rangle$ is not 
populated initially. 

In the results reported here, mean (Hartree-Fock) field effects are not included. In %%@
principle, shifts in the spectra occur if the atoms in the initial and final states of the %%@
RF-transfer, $|2\rangle$ and $|3\rangle$, feel a different mean field caused by atoms in %%@
state $|1\rangle$. We have analyzed the problem also assuming differing, density dependent %%@
mean fields for the states $|2\rangle$ and $|3\rangle$. In that case, a notable feature is %%@
that the free-atom peak position deviates from the bare-atom resonance 
position, also at temperatures well above $T_c$. In contrast, in the experiments ({\it %%@
31}), the free-atom peak is located at the bare-atom resonance position, displaying no %%@
mean field shift. Note that this is the case also at temperatures $\lesssim T_F$ where the %%@
free-atom peak is dominant and thus originates from atoms in the high density regions of %%@
the trap. The absence of such a mean field shift is related to the fact that, for $^6$Li, %%@
also the states $|1\rangle$ and $|3\rangle$ have a Feshbach resonance and unitarity %%@
limited interactions in close vicinity of the $|1\rangle$ -- $|2\rangle$ Feshbach %%@
resonance. Finally, the additional mean field shift that could be caused by the %%@
interaction between atoms in states $|2\rangle$ and $|3\rangle$ is absent due to the %%@
nature of the RF-field driven transition as was observed in ({\it 34}). 

Our method does not allow exact treatment of the onset of the pseudogap regime and precise 
study of the pseudogap transition temperature. However, the pseudogap pairing occurs at 
temperature $T^*$ slightly below the Fermi temperature $T_\mathrm{F}$ ({\it 35}). 
Our extrapolation scheme gives $T^* \approx 0.7 T_\mathrm{F}$. 
The Fermi temperature scales with the atom density as $T_\mathrm{F} \propto n^{(2/3)}$,
and the superfluid transition temperature follows approximately the same form at the 
unitarity limit. The free atom peak disappears at the temperature 
$T \sim 0.1\,T_\mathrm{F}$, which means that the temperature in terms of the local Fermi 
temperature of the border atoms is $T \sim 0.5\,T_\mathrm{F}^{0.1}$.
Here $T_\mathrm{F}^{0.1}$ is the local Fermi temperature scaled for density $n(r) = 0.1 %%@
n(0)$.
Therefore, the free-atom peak disappears when the border atoms are clearly below 
their (local) pseudogap temperature (however not yet below their critical temperature). 
In such temperatures, the atoms at the center of the trap, actually the majority of 
atoms, are already well below their local critical temperature. 

\subsection*{References and Notes} 

\begin{itemize}

\item[1.]
S.\ Jochim, M.\ Bartenstein, A.\ Altmeyer, G.\ Hendl, S.\ Riedl, C.\ Chin, J.\ Hecker 
Denschlag, R.\ Grimm, {\it Science} \textbf{302}, 2101 (2003).
\item[2.]
M.\ Greiner, C.\ A.\ Regal, D.\ S.\ Jin, {\it Nature} \textbf{426}, 537 (2003).
\item[3.]
M.\ W.\ Zwierlein, C.A.\ Stan, C.H.\ Schunck, S.M.F.\ Raupach, S.\ Gupta, Z.\ Hadzibabic, 
W.\ Ketterle, {\it Phys. Rev. Lett.} \textbf{91}, 250401 (2003).
\item[4.]
K.\ E.\ Strecker, G.\ B.\ Partridge, R.\ Hulet, {\it Phys.\ Rev.\ Lett.\ } \textbf{91}, %%@
080406 (2003).
\item[5.]
J.\ Cubizolles, T.\ Bourdel, S.\ J.\ J.\ M.\ F.\ Kokkelmans, G.\ V.\ Shlyapnikov, 
C.\ Salomon, {\it Phys. Rev. Lett.} \textbf{91}, 240401 (2003).
\item[6.]
C.\ A.\ Regal, M.\ Greiner, D.\ S.\ Jin, {\it Phys.\ Rev.\ Lett.\ } \textbf{92}, 040403 %%@
(2004).
\item[7.]
M.\ Bartenstein, A.\ Altmayer, S.\ Riedl, S.\ Jochim, C.\ Chin, J.\ Hecker Denschlag, and 
R.\ Grimm, {\it Phys.\ Rev.\ Lett.\ } \textbf{92}, 120401 (2004).
\item[8.]
M.\ W.\ Zwierlein, C.\ A.\ Stan, C.\ H.\ Schunck, S.\ M.\ F.\ Raupach, A.\ J.\ Kerman, 
W.\ Ketterle, {\it Phys.\ Rev.\ Lett.\ } \textbf{92}, 120403 (2004).
\item[9.]
J.\ Kinast, S.\ L.\ Hemmer, M.\ E.\ Gehm, A.\ Turlapov, J.\ E.\ Thomas, 
{\it Phys.\ Rev.\ Lett.\ } \textbf{92}, 150402 (2004).
\item[10.]
M.\  Bartenstein, A.\ Altmeyer, S.\ Riedl, S.\ Jochim, C.\ Chin, J.\ Hecker Denschlag, 
R.\ Grimm, {\it Phys.\ Rev.\ Lett.\ } \textbf{92}, 203201 (2004).
\item[11.]
T.\ Bourdel, L.\ Khaykovich, J.\ Cubizolles, J.\ Zhang, F.\ Chevy, M.\ Teichmann, 
L.\ Tarruell, S.\ J.\ J.\ M.\ F.\ Kokkelmans, C.\ Salomon, preprint available at 
http://arXiv.org/abs/cond-mat/0403091.
\item[12.]
%\bibitem{universal}
H.\ Heiselberg, {\it Phys.\ Rev.\ A} {\bf 63}, 043606 (2001).
\item[13.]
J. Carlson, S.-Y.\ Chang, V.\ R.\ Pandharipande, K.\ E.\ Schmidt, {\it Phys.\ Rev.\ Lett.\ %%@
} 
\textbf{91}, 50401 (2003). 
\item[14.]
G.\ M.\ Bruun, C.\ J. \ Pethick, {\it Phys.\ Rev.\ Lett.\ } \textbf{92}, 140404 (2004). 
\item[15.]
%\bibitem{uni4}
T.\ -L.\  Ho, {\it Phys.\ Rev.\ Lett.\ } \textbf{92}, 090402 (2004).
\item[16.]
S.\ Stringari, {\it Europhys.\ Lett.\ } \textbf{65}, 749 (2004).    
\item[17.]
M.\ Holland, S.\ J.\ J.\ M.\ F.\ Kokkelmans, M.\ L.\ Chiofalo, R.\ Walser, 
{\it Phys.\ Rev.\ Lett.\ } \textbf{87}, 120406 (2001).
\item[18.]
S.\ J.\ J.\ M.\ F.\ Kokkelmans, J.\ N.\ Milstein, M.\ L.\ Chiofalo, R.\ Walser, 
M.\ J.\ Holland, {\it Phys.\ Rev.\ A} \textbf{65}, 053617 (2002).
\item[19.]
E.\ Timmermans, K.\ Furuya, P.\ W.\ Milonni, A.\ K.\ Kerman, {\it Phys.\ Lett.\ A} 
\textbf{285}, 228 (2001).
\item[20.]
Y.\ Ohashi, A.\ Griffin, {Phys.\ Rev.\ Lett.\ } \textbf{89}, 130402 (2002).
\item[21.]
J.\ Stajic, J.\ M.\ Milstein, Q.\ Chen, M.\ L.\ Chiofalo, M.\ J.\ Holland, K.\ Levin, 
{\it Phys.\ Rev.\ A} \textbf{69}, 063610 (2004).
\item[22.]
G.\ M.\ Falco, R.\ A.\ Duine, H.\ T.\ C.\ Stoof, {\it Phys.\ Rev.\ Lett.\ } \textbf{92}, %%@
140402 (2004).
\item[23.]
H.\ P.\ B\"uchler, P.\ Zoller, W.\ Zwerger, preprint available at %%@
http://arXiv.org/abs/cond-mat/0404116.
\item[24.]
J.\ Stajic, A.\ Iyengar, Q.\ Chen, K.\ Levin, {\it Phys.\ Rev.\ B } \textbf{68}, 174517 %%@
(2003).
\item[25.]
J.\ Kinnunen, M.\ Rodr\'{\i}guez, P.\ T\"orm\"a, {\it Phys.\ Rev.\ Lett.\ } \textbf{92}, %%@
230403 (2004).
\item[26.]
A.\ Perali, P.\ Pieri, G.\ C.\ Strinati, preprint available at 
http://arXiv.org/abs/cond-mat/0405102.
\item[27.]
D.\ S.\ Petrov, C.\ Salomon, G.\ V.\ Shlyapnikov, preprint available at 
http://arXiv.org/abs/cond-mat/0309010.
\item[28.]
M.\ J.\ Holland, C.\ Menotti, L.\ Viverit, preprint available at 
http://arXiv.org/abs/cond-mat/0404234.
\item[29.]
M.\ Houbiers, R.\ Ferwerda, H.\ T.\ C.\ Stoof, W.\ I.\ McAlexander, C.\ A.\ Sackett, 
R.\ G.\ Hulet {\it Phys.\ Rev.\ A} \textbf{56}, 4864 (1997).
\item[30.]
M.\ L.\ Chiofalo, S.\ J.\ J.\ M.\ F.\ Kokkelmans, J.\ N.\ Milstein, M.\ J.\ Holland,
{\it Phys.\ Rev.\ Lett.\ } \textbf{88}, 090402 (2002). 
\item[31.]
C. Chin \emph{et al.}, Published online July 22 2004; 10.1126/science.1100818 (Science %%@
Express Reports).
\item[32.]
E.\ R.\ I.\ Abraham, W.\ I.\ McAlexander, J.\ M.\ Gerton, R.\ G.\ Hulet, R.\ Cot\'{e}, 
A.\ Dalgarno {\it Phys.\ Rev.\ A} \textbf{55}, 3299(R) (1997).
\item[33.]
We use a value for the Feshbach detuning that is slightly off from zero to the positive 
because, for the experiment in Fig.3 of ({\it 31}), a magnetic field corresponding to the 
upper error limit of the Feshbach resonance position was used.
\item[34.]
M.\ W.\ Zwierlein, Z.\ Hadzibabic, S.\ Gupta, W.\ Ketterle, {\it Phys.\ Rev.\ Lett.\ } 
\textbf{91}, 250404 (2003).
\item[35.]
C.\ A.\ R.\ S\'a de Melo, M.\ Randeria, J.\ Engelbrecht, {\it Phys.\ Rev.\ Lett.\ } 
\textbf{71}, 3202 (1993).

\end{itemize}

\end{document}